\documentclass[paper]{JHEP3}
\usepackage{amssymb,enumerate}
\usepackage{amsmath}
\usepackage{bbm}
\usepackage{url}
\usepackage{cite}
\usepackage{graphics}
\usepackage{xspace}
\usepackage{epsfig}
\usepackage{subfigure}

\newcommand\POWHEGBOX{{\tt POWHEG~BOX}}
\newcommand\PYTHIA{{\tt PYTHIA}}

\newcommand\HERWIG{{\tt HERWIG}}

\newcommand\vbfnlo{{\tt VBFNLO}}

\def\({\left(} 
\def\){\right)} 

\def\beq{\begin{equation}}
\def\beqn{\begin{eqnarray}}
\def\eeq{\end{equation}}
\def\eeqn{\end{eqnarray}}

\def\mr{\mathrm}

\def\vbfww{VBF $W^+W^-jj$\;}
\def\wpm{W^+W^-}

\def\evmv{e^+\nu_e\mu^-\bar\nu_\mu}

\def\muf{\mu_\mr{F}}
\def\mur{\mu_\mr{R}}

\title{Electroweak W$^+$W$^-jj$ prodution at NLO in QCD matched with
  parton shower in the POWHEG-BOX} \vfill

\author{
  Barbara J\"ager \\
PRISMA Cluster of Excellence \&  
Institute of Physics, Johannes Gutenberg University, 55099 Mainz, Germany\\
  E-mail: \email{jaegerba@uni-mainz.de} }

\author{Giulia Zanderighi \\
Rudolf Peierls Centre for Theoretical Physics, 1 Keble Road, University of Oxford, UK\\
E-mail: \email{g.zanderighi1@physics.ox.ac.uk}

}


\abstract{We present an implementation of electroweak $W^+W^-jj$
  production at hadron colliders in the {\tt POWHEG} framework, a
  method that allows the interfacing of a next-to-leading order QCD
  calculation with parton shower Monte Carlo programs. We provide
  results for both, fully and semi-leptonic decay modes of the weak
  bosons, taking resonant and non-resonant contributions and spin
  correlations of the final-state particles into account. To
  illustrate the versatility of our implementation, we provide
  phenomenological results for two representative scenarios with a
  light and with a heavy Higgs boson, respectively, and in a kinematic
  regime of highly boosted gauge bosons. The impact of the parton
  shower is found to depend on the setup and the observable under
  investigation. In particular, distributions related to a central-jet
  veto are more sensitive to these effects. Therefore the impact of
  radiation by the parton shower on next-to-leading order predictions
  should be assessed carefully on a case-by-case basis.  }
\preprint{MITP/13-004\\
  OUTP-13-005}

\begin{document}

\section{Introduction}
With the discovery of a new particle that is compatible with the
postulated Higgs boson at the CERN Large Hadron Collider
(LHC)~\cite{atlas:2012gk,cms:2012gu} and evidence for its existence at
the Fermilab Tevatron~\cite{Aaltonen:2012qt}, high-energy physics has
entered a new era. To clarify whether this particle with a mass of
about 125~GeV indeed is the CP-even, spin-zero Higgs boson predicted
by the Standard Model (SM), a determination of its properties, such as
its couplings to gauge bosons and fermions, spin and CP properties,
and decay width, is
indispensable~\cite{Zeppenfeld:2000td,Duhrssen:2004cv,LHCHiggsCrossSectionWorkingGroup:2012nn}.

In this context, electroweak vector boson fusion (VBF) processes play
a crucial
role~\cite{Rainwater:1999sd,Kauer:2000hi,Rainwater:1997dg,Rainwater:1998kj}. Higgs
production via VBF mainly proceeds via the scattering of quarks by the
exchange of weak gauge bosons in the $t$-channel that subsequently
radiate a Higgs boson. Because of the color singlet nature of the weak
gauge boson exchange, gluon radiation in the central-rapidity region
is strongly suppressed. The scattered quarks typically give rise to
two well-separated jets in the forward regions of the detector, while
the decay products of the Higgs boson tend to be located at central
rapidities, in between the two tagging jets. These characteristic
features of VBF reactions help to distinguish them from a priori
overwhelming QCD backgrounds.
Higgs production via VBF has been considered in the $H\to
\gamma\gamma$, $H\to \tau^+\tau^-$, and the fully leptonic $H\to
WW^{(*)}$ decay modes in the most recent analyses of the
ATLAS~\cite{atlas:2012gk} and CMS
collaborations~\cite{cms:2012gu}. Clearly, a precise knowledge of each
signal and background process is essential for quantitative results,
in particular if one aims at performing coupling measurements.

In this article, we focus on electroweak $\wpm$ production in
association with two tagging jets, $pp\to \wpm jj$. This process
contains both, the signal-type contributions from the VBF-induced
production of a Higgs boson that subsequently decays into a pair of
weak bosons, and the irreducible background from the continuum $\wpm$
production via VBF. To maintain unitarity, both contributions to the
full $\wpm jj$ final state at order $\mathcal{O}(\alpha^4)$ have to be
taken into account, even though in experimental analyses selection
cuts may be imposed to diminish the impact of the unwanted background
coming from the $\wpm$ continuum.  In order to best tune the selection
cuts, it is then clearly desirable to be able to simulate the
VBF-induced $\wpm jj$ production at the Higgs resonance as well as in
the $\wpm$ continuum in a common setup, taking both types of
contributions consistently into account at the highest level of
precision that is currently attainable for this class of reactions
\cite{Jager:2006zc,Jager:2006cp,Bozzi:2007ur,Jager:2009xx,Denner:2012dz,Jager:2011ms}. Accomplishing
this goal is the major purpose of this work. Building on existing
next-to-leading order (NLO) QCD calculations for weak boson pair
production via VBF~\cite{Jager:2006zc}, we aim at providing an
interface between the NLO-QCD calculation and parton shower programs
such as \HERWIG{}~\cite{Marchesini:1991ch,Corcella:2000bw} or
\PYTHIA{}~\cite{Sjostrand:2006za} to allow for precise, yet realistic
and flexible simulations of $\wpm jj$ production processes at hadron
colliders. To this end we develop an implementation of electroweak
$\wpm jj$ production in the context of the
\POWHEGBOX{}~\cite{Alioli:2010xd}, a framework for matching dedicated
NLO-QCD calculations with public parton-shower programs
\cite{Nason:2004rx,Frixione:2007vw}. We consider both the
fully-leptonic and the semi-leptonic decay modes of the $W$ bosons.

We describe the technical details of our implementation in
Sec.~\ref{sec:tech}. Section~\ref{sec:pheno} contains sample
phenomenological results for a few representative setups, where the
$W$ bosons decay either fully leptonically or semi-leptonically. In
the last case, we consider also a kinematical regime where the $W$
bosons are produced highly boosted.  The code has been made available
in the public repository of the \POWHEGBOX{} at the web site {\tt
  http://powhegbox.mib.infn.it}. Our conclusions are given in
Sec.~\ref{sec:conc}.

\section{Technical details}
\label{sec:tech}

\subsection{Next-to-leading order QCD corrections to \vbfww production}
The calculation of the next-to-leading order (NLO) QCD corrections to
\vbfww{} production with fully leptonic decays has first been
accomplished in Ref.~\cite{Jager:2006zc} and is publicly available in
the framework of the \vbfnlo~package~\cite{Arnold:2008rz}. We
extracted the matrix elements at Born level, the real emission and the
virtual corrections from that reference and adapted them to the format
required by the \POWHEGBOX. In addition, we are providing matrix
elements for the semi-leptonic decays of the weak bosons.

Electroweak $\wpm\,jj$ production with fully leptonic decays in
hadronic collisions mainly proceeds via the scattering of two
(anti-)quarks by the exchange of weak bosons in the $t$-channel, which
in turn emit $W$~bosons that may decay into lepton-neutrino pairs,
c.f.~Fig.~\ref{fig:feynman-tree}~(a). Furthermore, diagrams with one
or both of the weak bosons being emitted off a quark line occur,
c.~f.~Fig.~\ref{fig:feynman-tree}~(b), as well as $t$-channel
configurations with non-resonant $\ell^+\nu_\ell
{\ell'}^-\bar\nu_{\ell'} $ production in addition to the two jets,
c.~f.~Fig.~\ref{fig:feynman-tree}~(c).
%
%
\FIGURE[t]{
\includegraphics[angle=0,scale=0.7]{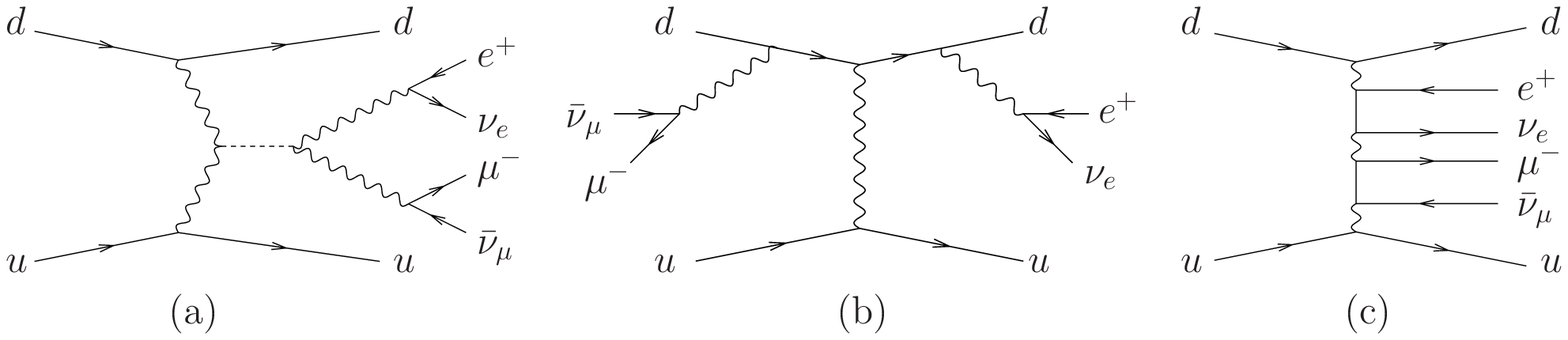}
\caption{Resonant (a,b) and non-resonant (c) sample diagrams for the
  partonic subprocess $du\to \evmv\, du$ at leading order.}
\label{fig:feynman-tree}
} 
%
In principle, the same final state may arise from quark-antiquark
annihilation diagrams with weak-boson exchange in the $s$-channel and
subsequent decay of one of the gauge bosons into a pair of jets. Such
contributions have been shown to be
negligible~\cite{Ciccolini:2007ec,Denner:2012dz} in regions of phase
space where VBF processes are searched for experimentally, and will
therefore be disregarded throughout. For subprocesses with identical
quark-flavor combinations, in addition to the $t$-channel topologies
one encounters $u$-channel exchange diagrams, which we include in our
calculation. Their interference with the $t$-channel diagrams is
however strongly suppressed once typical VBF selection cuts are
applied, and will therefore be neglected here.

In the following, we will refer to the electroweak production process
$pp\to \evmv\, jj$ within the above-mentioned approximations as
``fully leptonic \vbfww'' production, even though we include
contributions from non-resonant diagrams that do not arise from the
decay of a $\wpm\,jj$ intermediate state.
Final states related to one of the weak bosons in $pp\to \wpm jj$
decaying into a massless quark-antiquark and the other one into a
lepton-neutrino pair are referred to as ``semi-leptonic''. In analogy
to the fully leptonic decay mode, for semi-leptonic \vbfww production
we take double-, single- and non-resonant diagrams into account and
neglect $s$-channel contributions as well as interference effects
between identical quarks of the VBF production process.  In addition,
we disregard interference effects between the decay quarks and the
quarks of the VBF production process. Decays into massive quarks are
not taken into account. For a recent reference on the impact of quark
masses in semi-leptonic $H\to W W^\star$ decays, see, e.g.,
Ref.~\cite{Groote:2013xt}.  Representative diagrams for the partonic
subprocess $du \to c \bar s \mu^-\bar\nu_\mu \, du$ can be obtained by
replacing the $e^+\nu_e$ in Fig.~\ref{fig:feynman-tree} with $c \bar
s$ pairs. Similarly, some diagrams for the partonic subprocess $du \to
s \bar c \mu^-\bar\nu_\mu \, du$ are obtained by replacing the
$\mu^-\bar\nu_\mu$ in Fig.~\ref{fig:feynman-tree} with $s \bar c$
pairs.  We note, however, that some additional diagrams that, because
of the absence of a photon-neutrino coupling, do not occur for the
$\evmv\,jj$ final state have to be computed for each of the
semi-leptonic decay modes.

The NLO-QCD corrections to fully leptonic \vbfww production comprise
real-emission contributions with one extra parton in the final state
and the interference of one-loop diagrams with the Born
amplitude. Within our approximations, for the latter only self-energy,
triangle, box, and pentagon corrections to either the upper or the
lower fermion line have to be considered. The finite parts of these
contributions are evaluated numerically by means of a
Passarino-Veltman-type tensor reduction that is stabilized by means of
the methods of Refs.~\cite{Denner:2002ii,Denner:2005nn}. The numerical
stability is monitored by checking Ward identities at every
phase-space point. The real-emission contributions are obtained by
attaching a gluon in all possible ways to the tree-level diagrams
discussed above. Crossing-related diagrams with a gluon in the initial
state are also taken into account. Infrared singularities emerging in
both, real-emission and virtual contributions, at intermediate steps
of the calculation, are taken care of via the Frixione-Kunszt-Signer
subtraction approach~\cite{Frixione:1995ms} that is provided by the
\POWHEGBOX{} framework.
For semi-leptonic \vbfww production, QCD corrections to the $W$
hadronic decay are implemented only in the shower approximation. We
remark, however, that most shower Monte Carlo programs describe the
dressing of hadronic $W$ decays with QCD radiation accurately, since
they have been fit to LEP2 data.  The NLO-QCD corrections are thus of
the same form as those for the fully leptonic final state, but
hadronic decays include the inclusive ${\cal O}(\alpha_s)$ QCD
correction to the decay vertex.

\subsection{The \POWHEGBOX{} implementation}
For the implementation of \vbfww production in the framework of the
\POWHEGBOX{} we proceed in analogy to
Refs.~\cite{Jager:2011ms,Jager:2012xk}. Because of the larger
complexity of the current process, however, some further developments
are necessary.

As an input, the \POWHEGBOX{} requires a list of all independent
flavor structures of the Born and the real emission processes, the
Born amplitude squared, the real-emission amplitude squared, the
finite parts of the virtual amplitudes interfered with the Born, the
spin- and the color-correlated Born amplitudes squared. Because of the
simple color structure of VBF processes the latter are just multiples
of the Born amplitude itself, while the spin-correlated amplitudes
vanish entirely. For the fully leptonic decay modes, the LO amplitudes
squared, the virtual and the real-emission contributions are extracted
from Ref.~\cite{Jager:2006zc}, as described in the previous
section. For semi-leptonic decays of the weak bosons, appropriate
modifications of the matrix elements are performed.
Subtraction terms do not need to be provided explicitly, but are
computed by the \POWHEGBOX{} internally. Due to the special color
structure of VBF processes, within our approximations there is no
interference between radiation off the upper and the lower fermion
lines. This information has to be passed to the \POWHEGBOX{} by
assigning a tag to each quark line, as explained in some detail in
Refs.~\cite{Nason:2009ai,Jager:2011ms}. The tags are taken into
account by the \POWHEGBOX{} when singular regions for the generation
of radiation are identified.

Similarly to the case of electroweak $Zjj$
production~\cite{Jager:2012xk}, in the Born cross section for
electroweak $\wpm jj$ production collinear $q\to q\gamma$
configurations can arise when a $t$-channel photon of low virtuality
is exchanged. Such contributions are considered to be part of the QCD
corrections to $p\gamma\to \wpm\,jj$ and not taken into account here.
To effectively remove such contributions even before VBF cuts are
applied, we introduce a cut-off variable
$Q^2_{\gamma,\mr{min}}=4$~GeV$^2$ for the virtuality of the
$t$-channel exchange boson. Contributions from configurations with a
virtuality $Q^2$ below this cutoff value are dropped prior to
phase-space integration. We checked that, within the numerical
accuracy of the program, predictions do not change when the cut-off
variable is increased to 10~GeV$^2$.

To improve the efficiency of the program, in addition we are employing
a so-called Born-suppression factor $F(\Phi_n)$ that vanishes whenever
a singular region of the Born phase space $\Phi_n$ is approached. In
the \POWHEGBOX{}, the underlying Born kinematics is then generated
according to a modified $\bar B$ function,
\beq
\bar B_\mr{supp} = \bar B(\Phi_n) F(\Phi_n)\,.
\eeq
For \vbfww production, at Born level, singular configurations related
to the exchange of a photon of low virtuality in the $t$-channel are
characterized by low transverse momentum of an outgoing parton. It is
therefore advantageous to apply a Born-suppression factor that damps
such configurations. Following the prescription of
Ref.~\cite{Jager:2012xk}, we are using
\beq
F(\Phi_n) = \left(\frac{p_{T,1}^2}{p_{T,1}^2+\Lambda^2}\right)^2
 	    \left(\frac{p_{T,2}^2}{p_{T,2}^2+\Lambda^2}\right)^2\,,
\eeq
with the $p_{T,i}$ denoting the transverse momenta of the two outgoing
partons of the underlying Born configuration, and $\Lambda = 10$~GeV.

%
Electroweak $\wpm$ production in association with two jets contains
contributions from VBF-induced Higgs production with subsequent decay
into a pair of weak bosons, and from continuum $\wpm$ production via
VBF. These two types of contributions populate different regions of
phase space: The Higgs resonance exhibits a pronounced peak where the
invariant mass of the decay leptons and neutrinos, $M_\mr{decay}$, is
equal to the Higgs mass, whereas the $WW$~continuum is distributed
over a broad range in $M_\mr{decay}$. In order to optimize the
efficiency of the phase space integration, in the case of a light
Higgs boson with a narrow width we have split our simulation into two
contributions, depending on the value of $M_\mr{decay}$ in the
underlying Born configuration. In the region
\beq
\label{eq:higgs-region}
m_H-n \cdot\Gamma_H < M_\mr{decay} < m_H+n \cdot \Gamma_H\,, \qquad n = 50\,, 
\eeq
our results are then dominated by the sharp Higgs resonance, whereas
for other values of $M_\mr{decay}$ results are fully dominated by the
broad $WW$~continuum.
Even though both phase space regions contain contributions from all
Feynman diagrams, for simplicity we refer to the region of
Eq.~(\ref{eq:higgs-region}) as ``Higgs resonance'' and to the
complementary region as ``$WW$~continuum''.

We note that, because of the presence of two neutrinos, the invariant
mass distribution cannot be fully reconstructed in experiment in the
fully leptonic decay mode. The separation we perform is of purely
technical nature and serves the only purpose of improving the
convergence of the phase space integration.
To obtain meaningful results for the full VBF~$\wpm jj$ final state,
the two contributions therefore have to be added, and results are
independent of the choice of $n$ in Eq.~\eqref{eq:higgs-region}.
Still, it is interesting to observe the differences of some
characteristic distributions in these two
regions. Figure~\ref{fig:phill-inccuts}
\FIGURE[t]{
\includegraphics[angle=0,width=0.65\textwidth]{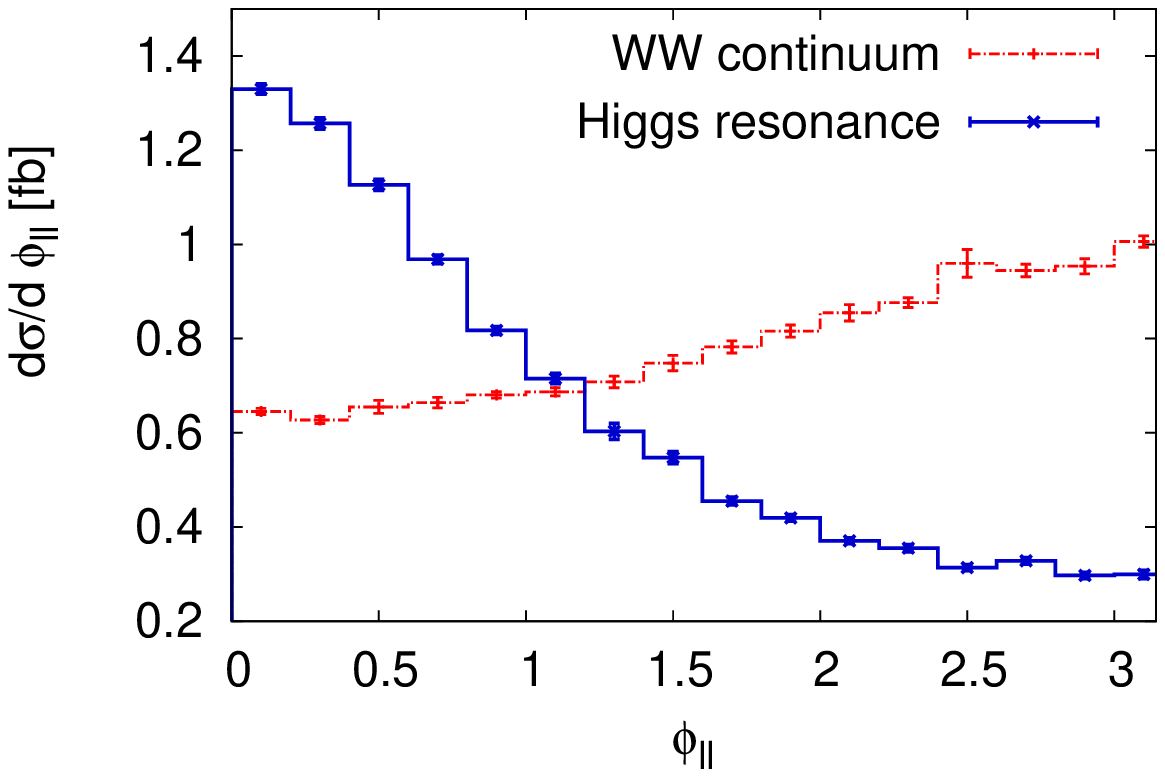}
\caption{Azimuthal angle separation of the two charged leptons at
  NLO-QCD accuracy for the Higgs resonance and the $\wpm$
  continuum region. See text for more details.
}
\label{fig:phill-inccuts}
} 
shows the azimuthal angle separation of the two charged leptons 
after only basic transverse momentum cuts of $p_{T,j}>25$~GeV
are applied on the two hardest jets that are reconstructed via the
anti-$k_T$ algorithm~\cite{Cacciari:2005hq,Cacciari:2008gp} in the
rapidity range $y_j<4.5$, with $R>0.4$.
As is evident, in the Higgs resonance region the leptons tend to be
close in azimuthal angle, as they arise from the decay of two weak
bosons that stem from a Higgs boson of spin zero. In the continuum
region, no such correlation exists between the decay leptons,
resulting in a completely different shape of the angular
distribution~\cite{Dittmar:1996ss}. Should one wish to enhance the
Higgs contributions with respect to the full \vbfww~cross section,
clearly one would make use of this feature.

%

In order to validate the complete implementation of \vbfww production
in the \POWHEGBOX{}, we have performed various checks.
The LO and real-emission matrix elements for each class of
subprocesses have been compared at the amplitude level to {\tt
  MadGraph}-generated
code~\cite{Stelzer:1994ta,Alwall:2007st}. 
We found agreement at the level of 12 significant digits.
With the user-supplied LO and real-emission matrix elements squared,
the \POWHEGBOX{} itself tests whether the real-emission cross section
approaches all soft and collinear limits correctly. This provides a
useful check on the relative normalization of the Born and
real-emission amplitudes squared as well as on the flavor summation.
For the fully leptonic decay mode, we have furthermore compared all
parts of the NLO-QCD calculation to the respective results generated
with the code of Ref.~\cite{Jager:2006zc}. We found full agreement for
integrated cross sections and differential distributions at LO and at
NLO-QCD, both within inclusive cuts and after imposing VBF-specific
selection criteria.

Finally, we remark that in order to produce the plots presented in
this work, we have used a new version of the \POWHEGBOX{} files that
were kindly provided to us by Paolo Nason and that will soon be
released as part of the Version 2 of the \POWHEGBOX{}. In particular
these files allow to compute the integration grids in parallel and
have a more efficient calculation of the upper bounds.

\section{Phenomenological results}
\label{sec:pheno}
Our implementation of \vbfww production in the \POWHEGBOX{} is
publicly available. Instructions for downloading the code are
available from the the web site of the \POWHEGBOX{} project, {\tt
  http://powhegbox.mib.infn.it}. Technical parameters of the code and
recommendations for its use can be found in a documentation that is
provided together with the code.
In this article, we present results obtained with our \POWHEGBOX{}
implementation for some representative setups. The user of the
\POWHEGBOX{} is of course free to perform studies with settings of her
own choice.

\subsection{Results at 8 TeV}
We consider proton-proton collisions at a center-of mass energy of
$\sqrt{s}=8$~TeV. For the parton distribution functions of the proton
we use the NLO set of the MSTW2008
parametrization~\cite{Martin:2009iq}, as implemented in the {\tt
  LHAPDF} library~\cite{Whalley:2005nh}. Jets are defined according to
the anti-$k_T$ algorithm~\cite{Cacciari:2005hq,Cacciari:2008gp} with
$R=0.4$, making use of the {\tt FASTJET}
package~\cite{Cacciari:2011ma}. As electroweak (EW) input parameters
we use the mass of the $Z$ boson, $m_Z=91.188$~GeV, the mass of the
$W$~boson, $m_W=80.419$~GeV, and the Fermi constant,
$G_F=1.16639\times 10^{-5}$~GeV$^{-1}$. Other EW parameters are obtain
from these via tree-level electroweak relations. For the widths of the
weak bosons we use $\Gamma_Z = 2.51$~GeV, $\Gamma_W=2.099$~GeV. In
case semi-leptonic decay modes are considered, the hadronic width is
corrected with a factor $\left[1+\alpha_s(m_W)/\pi\right]$. The
factorization and renormalization scales are set to $\muf=\mur=m_W$
throughout.

\subsubsection{Fully leptonic decay mode}
\label{sec:fully-lep}
\vbfww production with fully leptonic decays of the $W$~bosons is an
important channel in the Higgs search over a wide mass range. Here, we
present numerical results for electroweak $\evmv\,jj$ production at
the LHC within the setup outlined above. The mass of the Higgs boson
is set to $m_H=125$~GeV, the region where the {\tt ATLAS} and {\tt
  CMS} collaborations observe a new resonance compatible with the
Higgs boson predicted by the Standard Model
\cite{atlas:2012gk,cms:2012gu}. The width of the Higgs boson is set to
$\Gamma_H=0.00498$~GeV.
Our phenomenological study is inspired by the analysis strategy of
Ref.~\cite{ATLAS-HWW-2012}. We require the presence of two jets with
\beq
\label{eq:pttag-cuts}
p_{T,j}>25~\mr{GeV},\quad
y_j<4.5\,.
\eeq
The two hardest jets inside the considered rapidity range are referred
to as ``tagging jets''. These two tagging jets are furthermore
required to be well-separated from each other,
\beq
|y_{j1}-y_{j2}|<3.8\,,\quad
y_{j1}\times y_{j2} <0\,,\quad
m_{j1j2}>500~\mr{GeV}\,.
\eeq
We require missing energy and two hard charged leptons in the central
rapidity region,
\beq
y_\ell<2.5\,,\quad
p_{T,\ell_1}>25~\mr{GeV}\,,\quad
p_{T,\ell_2}>15~\mr{GeV}\,,\quad
p_{T}^\mr{miss}>25~\mr{GeV}\,,\quad
\eeq
which are well-separated from each other and from the jets,
\beq
R_{\ell\ell}>0.3\,,\quad
R_{j\ell}>0.3\,,
\eeq
but close in azimuthal angle, 
\beq
\label{eq:phill-cut}
|\phi_{\ell_1}-\phi_{\ell_2}|<1.8\,,
\eeq
and located in the rapidity region between the two tagging jets,
\beq
\label{eq:rap-gap}
\mr{min}\{y_{j1},y_{j2}\}<y_\ell<\mr{max}\{y_{j1},y_{j2}\}\,.
\eeq
When these cuts are applied, the inclusive cross section contributions
for VBF $\evmv\, jj$ production given in Sec.~\ref{sec:tech} are given
by $\sigma_{WW}^\mr{VBF}=(0.202\pm 0.002)$~fb and
$\sigma_\mr{Higgs}^\mr{VBF}=(0.268\pm 0.003)$~fb at NLO-QCD, amounting
to a total of $\sigma^\mr{VBF}=(0.470\pm 0.003)$~fb. Because of our
selection cuts, in particular the cut on the azimuthal angle
separation of the charged leptons, Eq.~({\ref{eq:phill-cut}), the
  Higgs contribution is slightly preferred compared to the
  $WW$~continuum. The VBF cross section at NLO QCD changes by less
  than 2\% when the factorization and renormalization scales are
  varied simultaneously in the range $\muf=\mur=m_W/2$ to $2
  M_W$. Slightly larger scale uncertainties are found for
  distributions related to jets that are emitted in addition to the
  two tagging jets. As such jets can first occur via the real-emission
  contributions of the NLO calculation, their kinematic properties are
  effectively described at LO only, and thus plagued by larger scale
  uncertainties than true NLO observables. While our calculation has
  been performed for a fixed choice of scale that should facilitate
  comparison to future calculations, we mention that more
  sophisticated choices are possible with our code, for instance like
  those suggested in the context of the MINLO
  method~\cite{Hamilton:2012np}.

  Since the invariant mass of the $\evmv$ system cannot be fully
  reconstructed, it is common to consider the transverse mass instead,
  which is defined as
\beq
\label{eq:mtww}
m_{T, WW} = \sqrt{
(E_T^{\,\ell\ell}+E_T^\mr{\,miss})^2-
|\vec{p}_T^{\;\ell\ell}+\vec{p}_T^\mr{\;miss}|^2
}\,,\quad
\text{with }
E_T^{\ell\ell} = \sqrt{
|\vec{p}_T^{\;\ell\ell}|^2+m_{\ell\ell}^2
}\,.
\eeq
Figure~\ref{fig:mtww-vbfcuts}~(a) 
%
%
%
\FIGURE[t]{
\includegraphics[angle=0,width=0.45\textwidth]{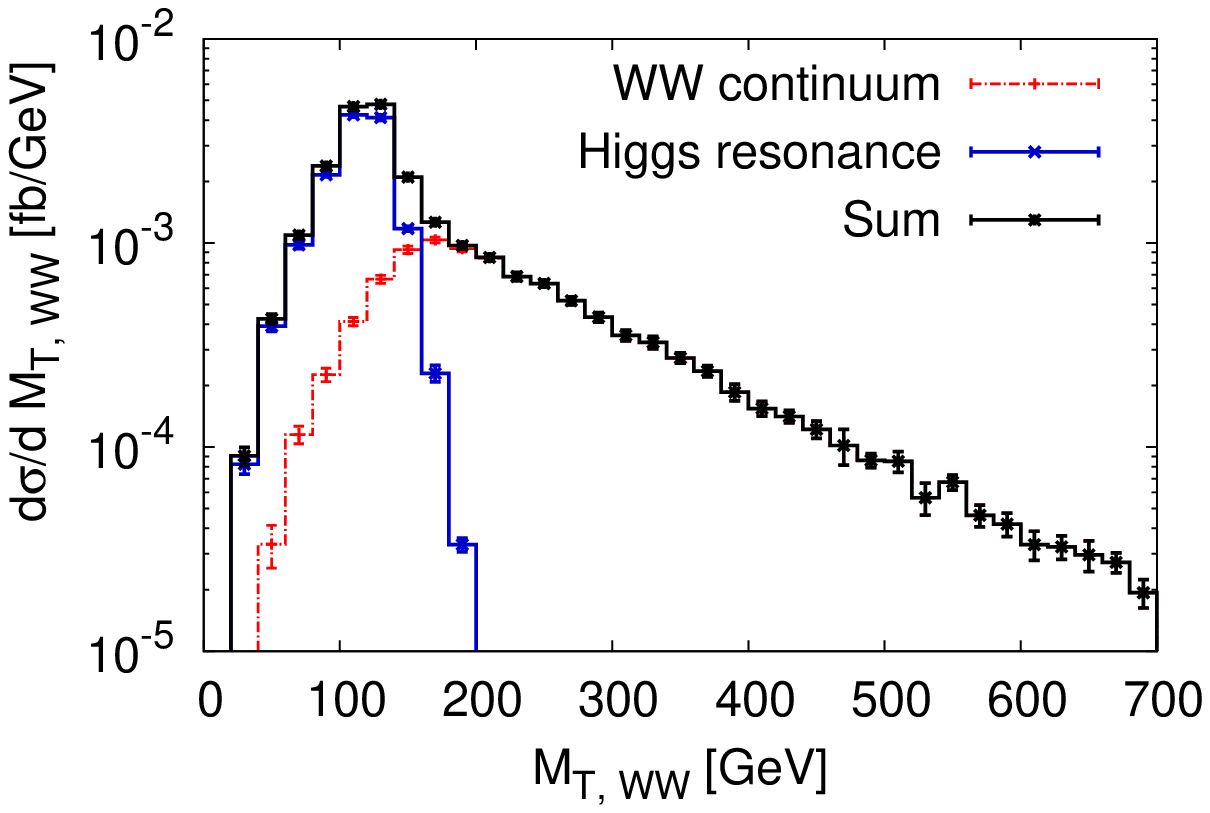}
\includegraphics[angle=0,width=0.45\textwidth]{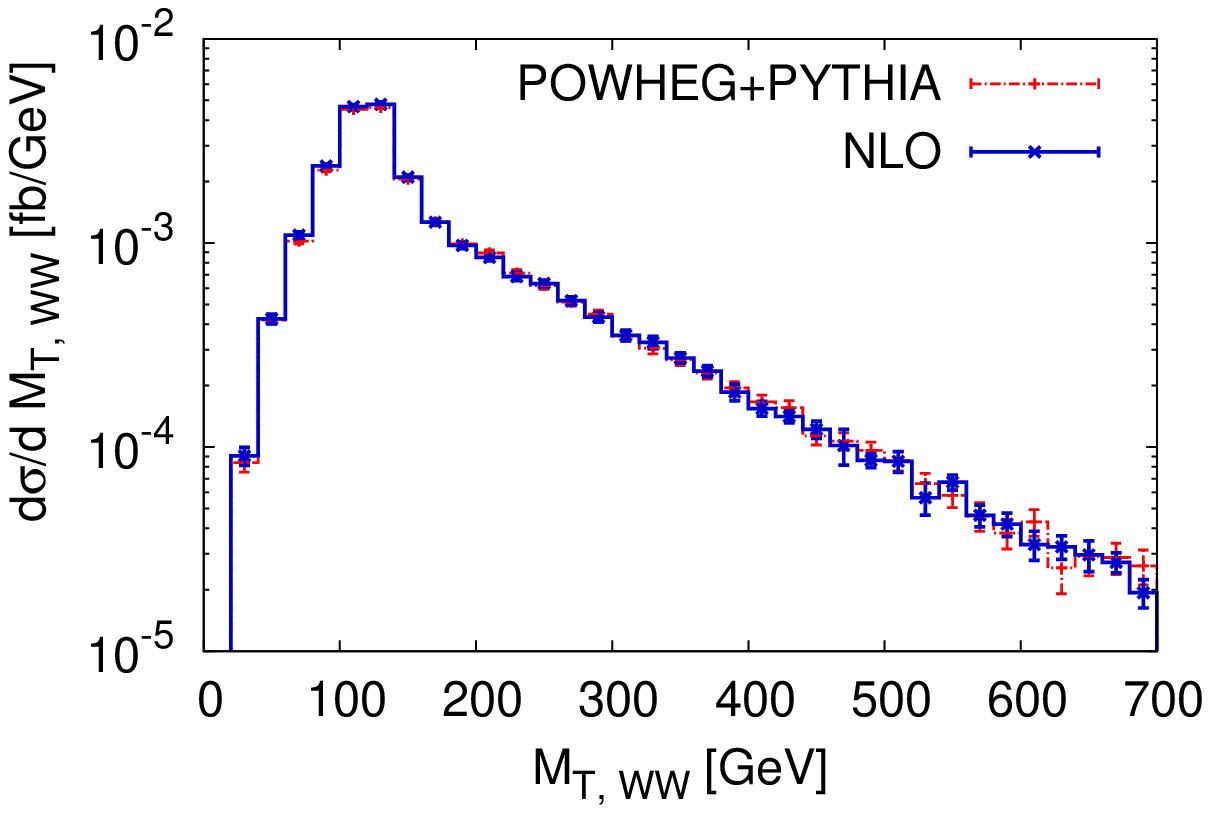}
\caption{Transverse mass distribution of the four-lepton system, as
  defined in Eq.~(\ref{eq:mtww}), for $pp\to\evmv\,jj$ at the LHC with
  $\sqrt{s}=8$~TeV within the VBF cuts of
  Eqs.~(\ref{eq:pttag-cuts})--(\ref{eq:rap-gap}). Left panel: results
  at NLO-QCD accuracy for the Higgs resonance~(blue line), the
  $WW$~continuum~(red line), and their sum~(black line). Right panel:
  Sum of the two contributions at NLO-QCD~(blue solid lines) and with {\tt
    POWHEG+PYTHIA}~(red dashed lines).}
\label{fig:mtww-vbfcuts}
} 
%
%
shows the respective contributions to $d\sigma^\mr{VBF}/dm_{T, WW}$ of
the Higgs resonance and of the $WW$~continuum, as defined in
Sec.~\ref{sec:tech}, as well as their sum. The Higgs contribution is
peaked at $m_{T, WW}\sim m_H$, while the $WW$~continuum is largest in
the kinematic range where two weak bosons can be produced on-shell.

In Fig.~\ref{fig:mtww-vbfcuts}~(b), we show the transverse mass
distribution at pure NLO QCD, as well as at NLO matched with a
parton-shower program via {\tt POWHEG} ({\tt NLO+PS}). For the parton
shower, here and in the following we are using
\PYTHIA{}~6.4.25~\cite{Sjostrand:2006za} with the Perugia 0 tune for
the shower, including hadronization corrections, multi-parton
interactions and underlying event. We do not take QED radiation
effects into account.  The shape of the transverse mass distribution
is only barely affected by parton-shower effects.  Similarly small
distortions of shapes are observed for the transverse momentum and
rapidity distributions of the leptons, as well as for their azimuthal
angle separation.
The overall normalization of the VBF cross sections
decreases by about 2\% when the NLO calculation is combined with {\tt
  PYTHIA}. 
The parton shower mostly gives rise to the emission of soft or
collinear radiation, while the probability for the emission of {\em
  hard} extra jets does not increase because of parton-shower
effects. Indeed, the production rate of a third hard jet decreases
when the NLO calculation is merged with {\tt PYTHIA}, as illustrated
in Fig.~\ref{fig:ptj3-vbfcuts},
%
%
\FIGURE[t]{
\includegraphics[angle=0,width=0.7\textwidth]{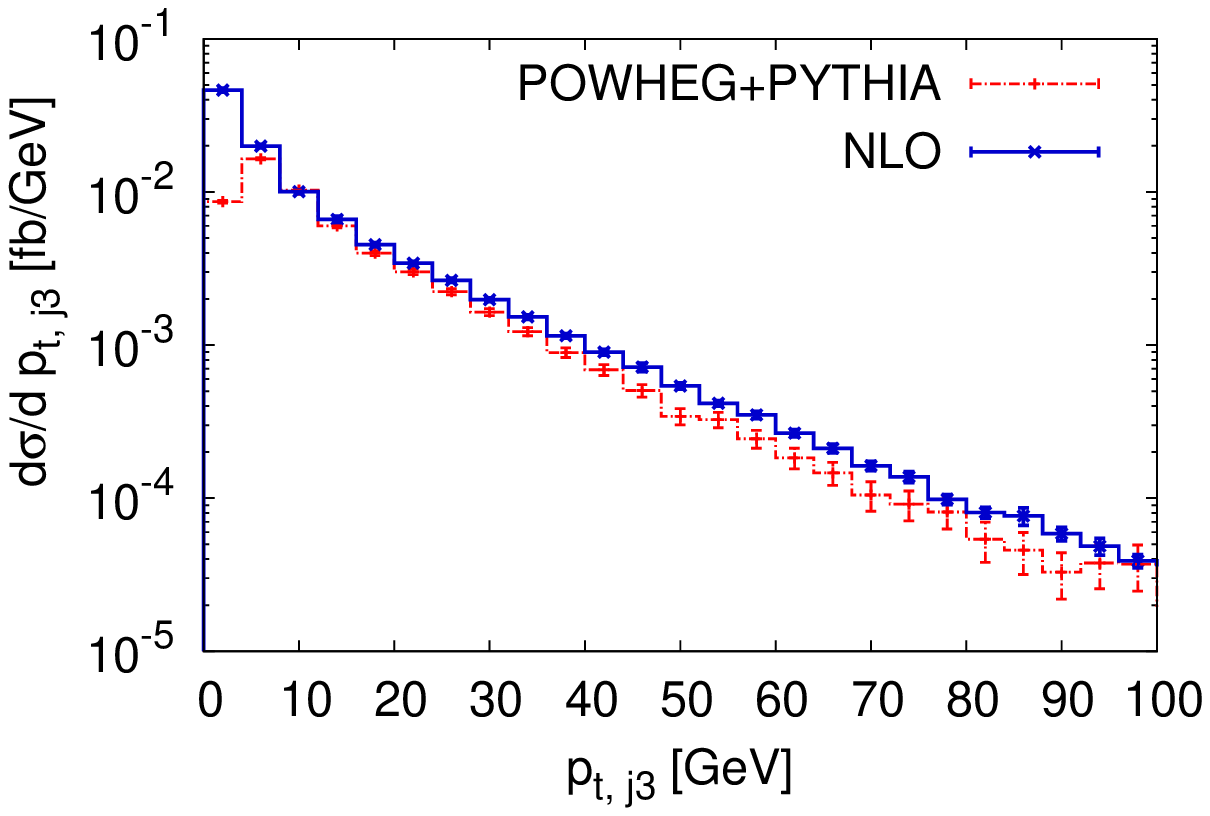}
\caption{ Transverse momentum distribution of the third jet in
  $pp\to\evmv\,jj$ at the LHC within the VBF cuts of
  Eqs.~(\ref{eq:pttag-cuts})--(\ref{eq:rap-gap}) at NLO-QCD (blue solid
  lines) and with {\tt POWHEG+PYTHIA}~(red dashed lines). }
\label{fig:ptj3-vbfcuts}
} 
%
%
which shows the transverse momentum distribution of the third jet in
$pp\to\evmv\,jj$ at NLO~QCD and for {\tt POWHEG+PYTHIA}. At NLO, only
the real-emission contributions can give rise to a third
jet. Distributions related to this jet are thus effectively described
only at the lowest non-vanishing order in the fixed-order predictions,
while in the {\tt POWHEG+PYTHIA} results soft-collinear radiation is
resummed at leading-logarithmic accuracy via the Sudakov factor,
resulting in a damping of contributions with very small $p_{T,j_3}$.

A quantitative understanding of central jets that are located in
between the two tagging jets is of crucial importance for the
discrimination of VBF events from hard QCD processes as well as from
underlying event and pile-up effects at the LHC. These backgrounds are
characterized by a considerable amount of jet activity in the
central-rapidity region of the detector, whereas the emission of hard
jets at central rapidities is strongly suppressed in VBF processes.
Figure~\ref{fig:y3-vbfcuts}
%
%
%
\FIGURE[t]{
  \includegraphics[angle=0,width=0.45\textwidth]{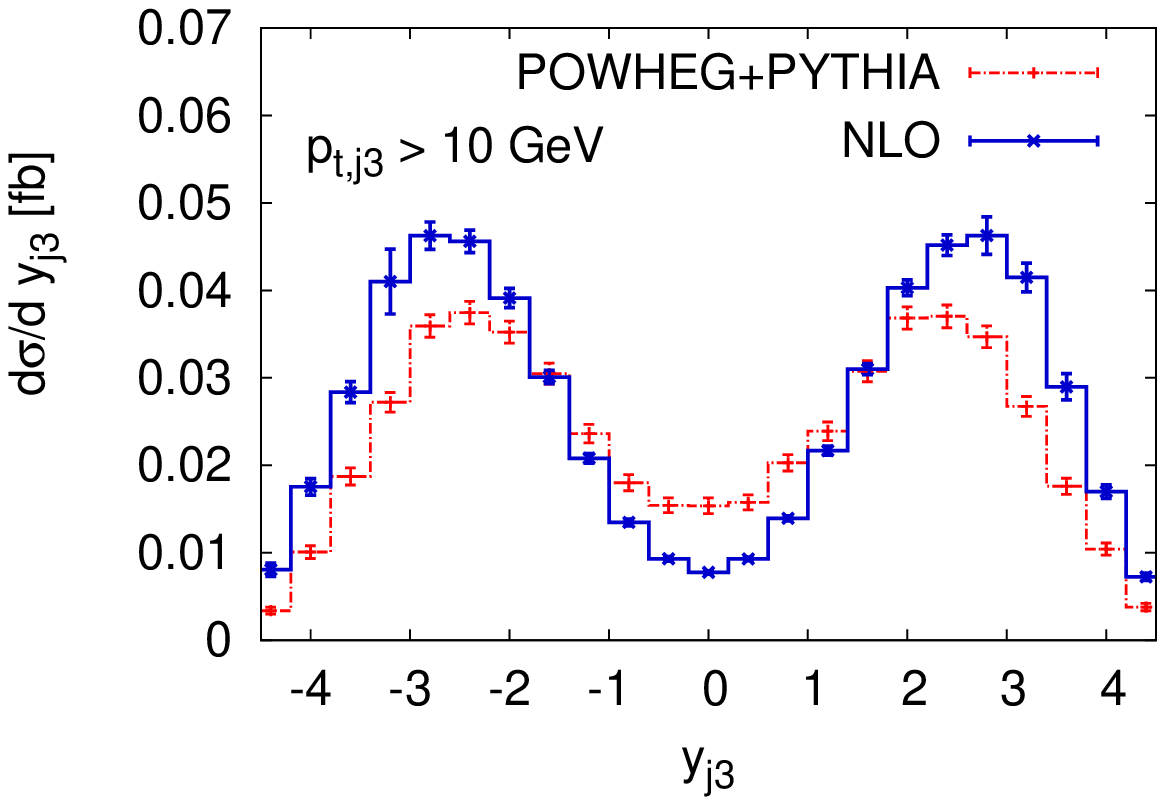}
  \includegraphics[angle=0,width=0.45\textwidth]{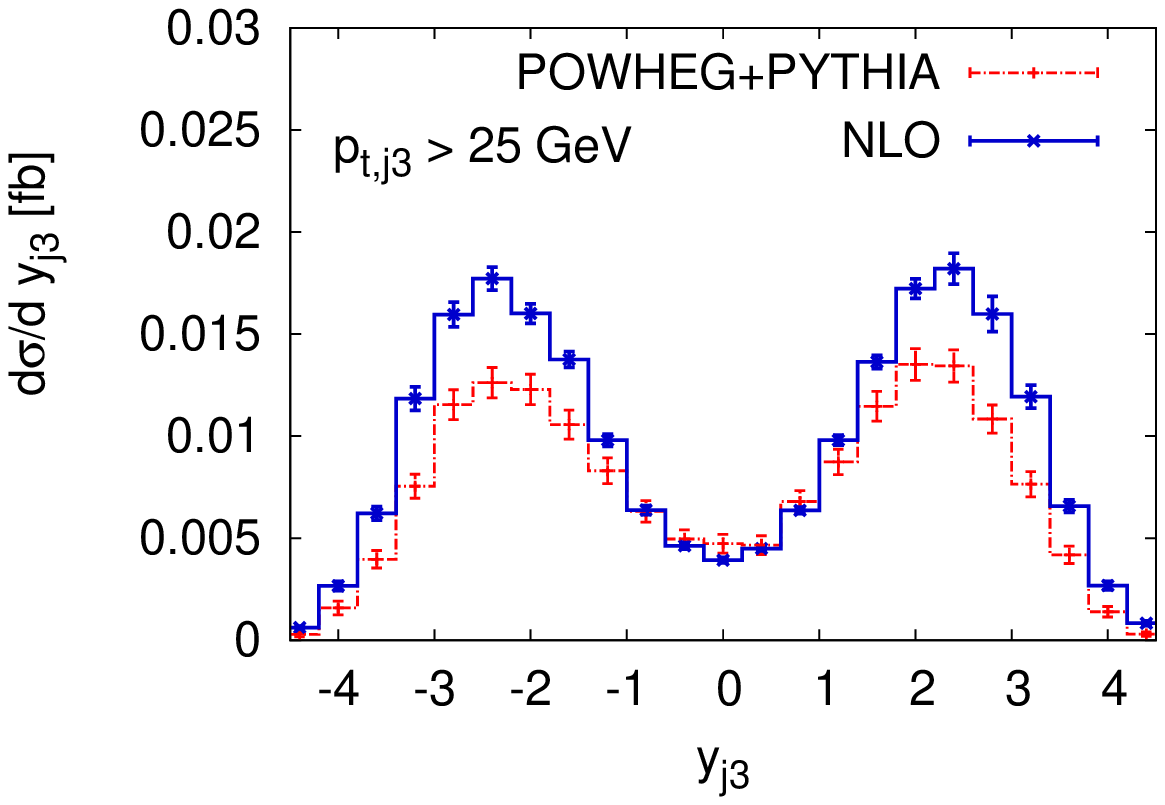}
  \caption{Rapidity distribution of the third jet in
    $pp\to\evmv\,jj$ at the LHC within the VBF cuts of
    Eqs.~(\ref{eq:pttag-cuts})--(\ref{eq:rap-gap}) and an additional
    cut of $p_{T,j_3}>10$~GeV~(left panel) and
    $p_{T,j_3}>25$~GeV~(right panel), respectively, at NLO-QCD (blue solid
    lines) and with {\tt POWHEG+PYTHIA}~(red dashed lines). }
\label{fig:y3-vbfcuts}
} 
%
shows the rapidity distribution of the third jet in VBF-induced
$\evmv\,jj$ production at the LHC. Obviously, the shape of the NLO-QCD
distributions is not particularly sensitive to the transverse momentum
cut imposed on the third jet. However, when the NLO calculation is
merged with {\tt PYTHIA}, the central-rapidity region is considerably
filled by extra jets, as illustrated in the left panel of
Fig.~\ref{fig:y3-vbfcuts}, where in addition to the VBF cuts of
Eqs.~(\ref{eq:pttag-cuts})--(\ref{eq:rap-gap}) we impose a loose cut
of $p_{T,j_3}>10$~GeV on the third-hardest jet. A significant
reduction of the central-jet activity can be achieved by tightening
this cut to $p_{T,j_3}>25$~GeV, as shown in the right panel of the
same figure.

The relative position of the third jet with respect to the two tagging
jets is accessible via the variable
\beq
y^\star = y_{j_3} - \frac{y_{j_1} + y_{j_2} }{2}\,. 
\eeq
In Fig.~\ref{fig:ystar-vbfcuts}~(a)  
%
%
%
\FIGURE[t]{
  \includegraphics[angle=0,width=0.45\textwidth]{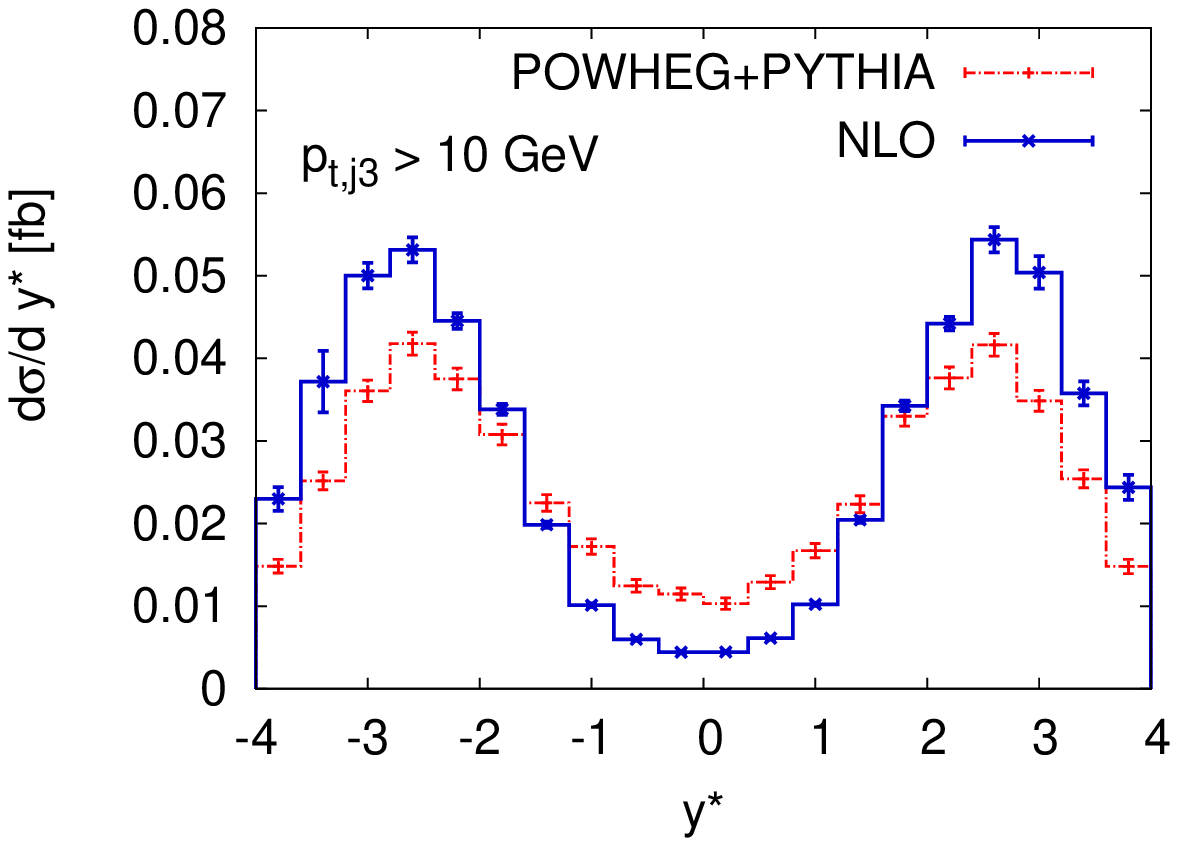}
  \includegraphics[angle=0,width=0.45\textwidth]{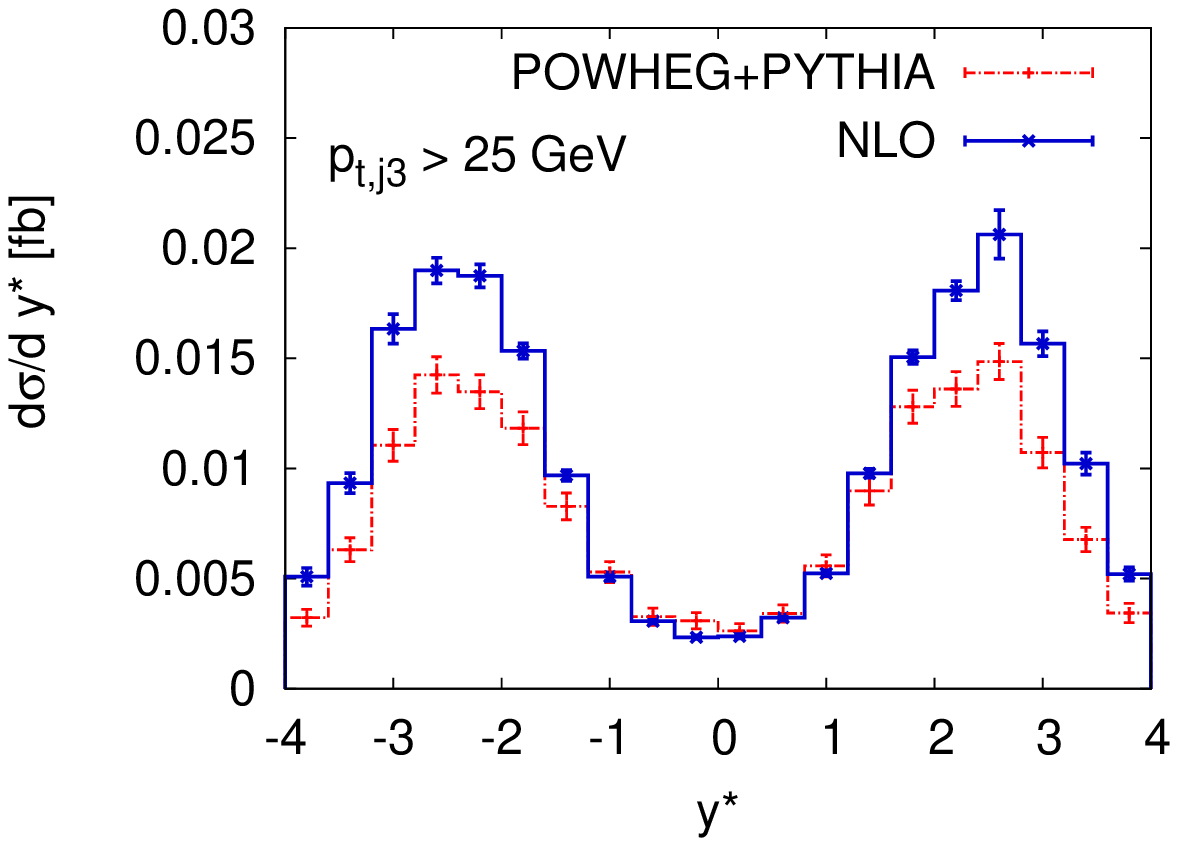}

  \caption{Rapidity distribution of the third jet with
    respect to the average of the two tagging jets in $pp\to\evmv\,jj$
    at the LHC within the VBF cuts of
    Eqs.~(\ref{eq:pttag-cuts})--(\ref{eq:rap-gap}), and an additional
    cut of $p_{T,j_3}>10$~GeV~(left panel) and
    $p_{T,j_3}>25$~GeV~(right panel), respectively, at NLO-QCD (blue solid
    lines) and with {\tt POWHEG+PYTHIA}~(red dashed lines). }
\label{fig:ystar-vbfcuts}
} 
%
%
the $y^\star$ distribution 
is shown within the VBF cuts of
Eqs.~(\ref{eq:pttag-cuts})--(\ref{eq:rap-gap}) and with two different
cuts on the third jet, $p_{T,j_3}>10$~GeV and $p_{T,j_3}>25$~GeV,
respectively. As for the rapidity distribution of the third jet, we
observe that the parton shower tends to fill the central
region. Increasing the transverse momentum cut of the third jet
reduces this central activity. 

Central-jet veto techniques
\cite{Barger:1994zq,Rainwater:1999sd,Kauer:2000hi} are based on the
observation that QCD backgrounds can be suppressed efficiently, if one
rejects events with one or more hard jets between the two tagging
jets, for instance, with
\beq
\label{eq:cjv}
p_T^\mr{j,veto}<20~\mr{GeV}\,,\quad
\mr{min}\{y_{j1},y_{j2}\}<y_\mr{j,veto}<\mr{max}\{y_{j1},y_{j2}\}\,.
\eeq
Because of the small amount of jet activity in the central region,
this central-jet veto diminishes the VBF signal only marginally,
resulting in a
%
decrease of about 12\% in the NLO and  {\tt NLO+PS} cross sections.

\subsubsection{Semi-leptonic decay mode}
\label{sec:semi-lep}
While fully leptonic decays of a Higgs boson produced via VBF give
rise to a rather clean final state with little jet activity in the
central-rapidity region, the invariant mass of the system consisting
of two charged leptons and two neutrinos cannot be fully
reconstructed. For Higgs masses above the $W$-pair production
threshold, the $H\to WW\to \ell\nu\,jj$ channel, with one of the weak bosons
decaying into a pair of jets, while the other one decays into a
lepton-neutrino pair, thus becomes interesting. In this channel, the
invariant mass of the decay system can be reconstructed, using
kinematical constraints to estimate the longitudinal component of the
neutrino momentum. A recent search for a Higgs boson in the range
$300\leq m_H\leq 600$~GeV by the ATLAS collaboration \cite{Aad:2012me}
is thus based on the $\ell\nu\,jj$ final state.

Here, we wish to show some representative results for \vbfww
production with semi-leptonic decays of the weak bosons in the
presence of one heavy Higgs boson, following closely the analysis
strategy of Ref.~\cite{Aad:2012me}. For the Higgs
boson we assume a mass of $m_H=400$~GeV and a width of
$\Gamma_H=26.791$~GeV.

We require the presence of at least four hard jets with transverse
momenta larger than 25~GeV. The decay jets of the hadronically
decaying $W$ boson are then identified as the pair of two out of the
four hardest jets whose invariant mass is closest to $m_W$. If we do
not find a pair of jets in the range
\beq
\label{eq:mdec-window}
71~\mr{GeV}\leq m_{jj}^\mr{dec}\leq 91~\mr{GeV}, 
\eeq
we discard the respective event. The remaining two hardest jets with 
\beq
p_{T,j}>25~\mr{GeV}, \;
y_j < 4.5\,,
\eeq
are identified as the tagging jets, which have to fulfill the
following requirements:
\beq
|y_{j1}-y_{j2}|<3\,,\quad
y_{j1}\times y_{j2} <0\,,\quad
m_{j1j2}>600~\mr{GeV}\,.
\eeq
In addition, we require a hard lepton and large missing energy,
\beq
p_{T,\ell}>30~\mr{GeV}, \;
p_{T}^\mr{miss}>30~\mr{GeV}. 
\eeq
The tagging jets have to be well-separated from the hard charged
lepton and from the decay jets,
\beq
R_{j\ell}>0.3,\,\quad 
R_{jd}>0.3,\,
\eeq
Both, the decay jets and the hard charged lepton have to be located in
the rapidity range between the two tagging jets,
\beq
\label{eq:rap-gap-lj}
\mr{min}\{y_{j1},y_{j2}\}<y_\ell, y_j^\mr{dec}<\mr{max}\{y_{j1},y_{j2}\}\,.
\eeq

In contrast to the fully leptonic decay modes, the invariant mass of
the decay system can be reconstructed, if the $W$~bosons are produced
on-shell. Following the prescription of Ref.~\cite{Aad:2012me}, for
each event the transverse component of the neutrino is identified with
the missing transverse energy. The longitudinal momentum of the
neutrino is computed via the quadratic equation given by the mass
constraint $m_{\ell\nu} = m_W$. In case of two real solutions, the
solution resulting in a smaller longitudinal neutrino momentum
component, $|p_{\nu}^z|$, is taken.  In case of complex solutions, the
event is disregarded.
Figure~\ref{fig:m2l2j} 
%
%
\FIGURE[t]{
  \includegraphics[angle=0,width=0.6\textwidth]{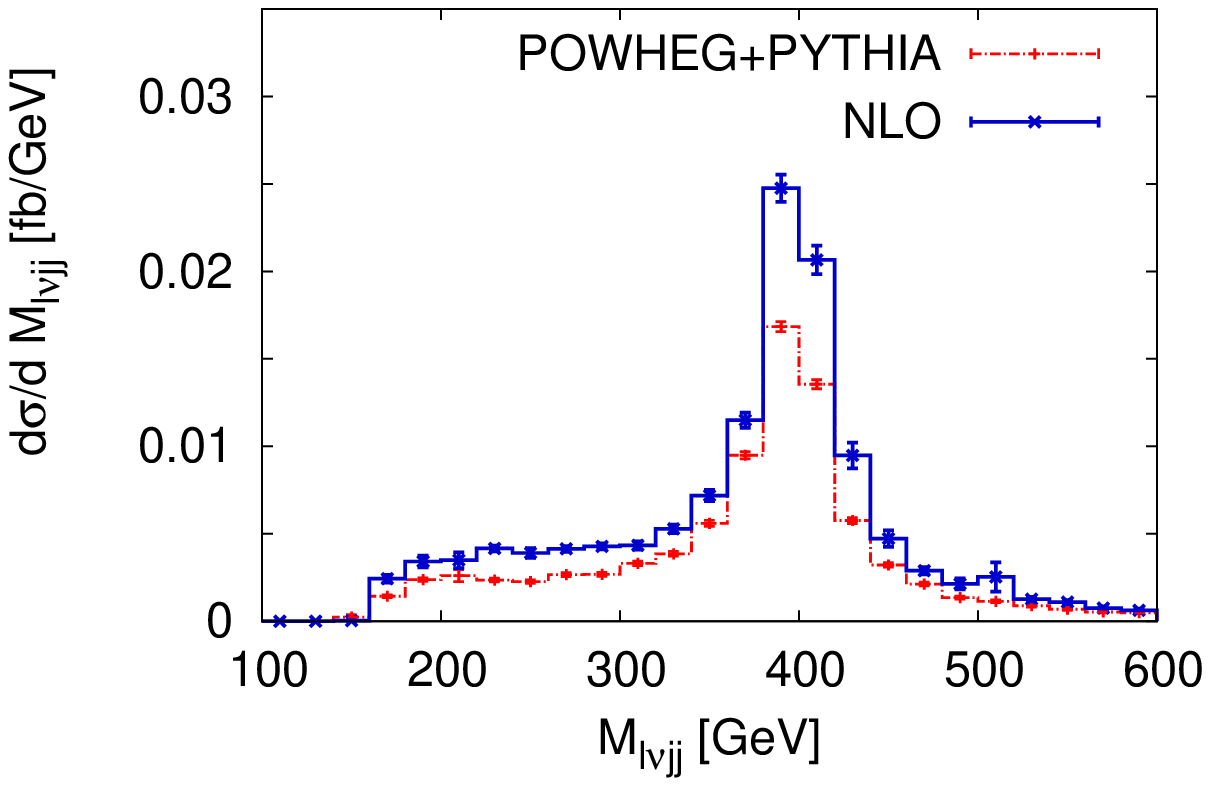}
  \caption{Invariant-mass distribution of the decay system in $pp\to
    W^+(jj) W^-(\mu^-\bar \nu_\mu)\,jj$ at the LHC within the VBF cuts of
    Eqs.~(\ref{eq:mdec-window})--(\ref{eq:rap-gap-lj}), at NLO-QCD
    (blue solid lines) and with {\tt POWHEG+PYTHIA}~(red dashed lines), see text for details.}
\label{fig:m2l2j}
} 
%
%
shows the invariant mass of the decay system, $M_{\ell\nu jj}$, with the neutrino
momentum reconstructed as described above. The peak of $M_{\ell\nu jj}$
at the value of the heavy Higgs boson's mass, $m_H=400$~GeV, is
clearly visible at NLO QCD. Its position is retained in the {\tt
  POWHEG+PYTHIA} result. However the normalization of the distribution
is modified considerably when the NLO calculation is merged with {\tt
  PYTHIA}. This effect can be traced back to a reduction of the cross
section for semi-leptonic \vbfww production within the above cuts, by
about 30\% when going from NLO to {\tt NLO+PS}.
%
This large difference can be explained by observing that invariant
mass distributions receive very large corrections from the parton
shower. Indeed, in the setup of ref.~\cite{Aad:2012me}, that we
adopted here, one requires the invariant mass of the two jets stemming
from the $W$ decay to be still close to the $W$ mass after parton
showering. However, the soft radiation from the parton shower tends to
smear this mass distribution, and, as a consequence, a large number of
events are discarded since they do not satisfy the condition in
Eq.~(\ref{eq:mdec-window}).
%
A way to avoid these large corrections is to consider a kinematical
regime where the $W$ bosons are highly boosted and decay into a single
fat jet. In this case, jet-substructure techniques can be used to
distinguish the jets coming from the $W$ decay from the other jets. We
outline such a study in the next section.


\subsection{Semi-leptonic decay mode with boosted kinematics}
At the LHC, in particular when it is running at its design energy of
$\sqrt{s}=14$~TeV (or close to it), heavy particles are copiously
produced with transverse momenta that greatly exceed their rest
mass. These boosted objects give rise to highly collimated streams of
decay particles in the detector.
It is then useful to exploit the feature that the partons formed in the
decay of the boosted object are close to each other, and, with a
suitable jet-algorithm, will end up in one single jet.  Modern jet
deconstruction techniques have been designed to resolve the
substructure of the so-called {\em fat jet} that arises from the
hadronic decay of the heavy, boosted object and to discriminate this
jet from ordinary QCD jets (see, e.g., Ref.~\cite{Abdesselam:2010pt}
for a recent review). 

We here present an analysis of electroweak $\wpm\,jj$ production,
assuming the presence of just a light Higgs boson with $m_H=125$~GeV,
in the kinematic regime that is particularly interesting in the
context of weak boson scattering.  Possible new physics in the gauge
boson sector is expected to manifest itself in the scattering of the
longitudinal gauge boson modes, $W_L W_L\to W_L W_L$, at high
energies. To access this region in \vbfww production with
semi-leptonic decays of the weak bosons, we closely follow the
strategy of Ref.~\cite{Butterworth:2002tt}, supplemented by the
jet-filtering techniques of Ref.~\cite{Butterworth:2008iy}. We require
the presence of a highly-boosted jet with
\beq
\label{eq:pt-boost}
p_{T,j}^\mr{boosted}>300~\text{GeV}\,,
\eeq
which is constructed with $R=1.0$ using the Cambridge/Aachen
algorithm~\cite{Dokshitzer:1997in,Wobisch:1998wt} for flexibly
resolving different angular scales. Similar to what is done in
Ref.~\cite{Butterworth:2008iy} we 
look for a boosted jet with 
invariant mass $M_J$ close to the mass of the $W$~boson,
\beq
m_W-10~\mr{GeV}<M_J<m_W+10~\mr{GeV}\,.
\eeq
To ensure that this fat jet is related to the decay of a heavy boosted
gauge boson, we additionally investigate its composition and demand
that it consists of two subjets $J_1,J_2$ with invariant masses
$m_{J_1}<m_{J_2}$. If there is a significant mass drop with respect to
the fat jet,
\beq
m_{J_1}<0.67\cdot M_J\,,\quad\text{and}\quad 
y = \frac{\mr{min}\,(p_{T,J_1}^2,p_{T,J_2}^2)}{M_J^2}\Delta R_{J_1,J_2} > 0.09\,,
\eeq
the boosted jet is deemed compatible with the $W$-decay. 
If no jet in the event satisfies these cuts, 
the event is discarded. 
The remaining two hardest jets with 
\beq
p_{T,j}>25~\mr{GeV}, \;
y_j < 4.5\,,
\eeq
are identified as the tagging jets, which have to fulfill the
following requirements:
\beq
|y_{j1}-y_{j2}|<3\,,\quad
y_{j1}\times y_{j2} <0\,,\quad
m_{j1j2}>600~\mr{GeV}\,.
\eeq
The hardest lepton also is forced into the boosted regime by the transverse momentum cut
\beq
p_{T,\ell}>300~\mr{GeV}\,.
\eeq
The longitudinal momentum component of the neutrino is computed in the
same way as in the analysis described in Sec.~\ref{sec:semi-lep}. 
In addition, we require 
\beq
p_T^\mr{miss}>30~\mr{GeV}\,,\quad
R_{j\ell}>0.3\,.
\eeq
with the latter denoting the $R$-separation between the tagging jets
$j$ and the lepton. 
The lepton and the fat jet need to be located in between the two tagging jets,
\beq
\label{eq:gap-boost}
\mr{min}\{y_{j1},y_{j2}\}<y_\ell, y_\mr{fat}<\mr{max}\{y_{j1},y_{j2}\}\,.
\eeq

Within this setup, rather small cross sections are obtained, amounting
to $\sigma^\mr{NLO}=(0.0458\pm 0.0006)$~fb at NLO.
Nonetheless, Ref.~\cite{Butterworth:2002tt} suggests that with
100~fb$^{-1}$ of luminosity at 14 TeV the LHC can distinguish between
$WW$ scattering in the Standard Model and in scenarios of new physics
with extra heavy scalar or vector resonances.
The invariant mass distribution of the $WW$ system is particularly
interesting in this context. We show this distribution in
Fig.~\ref{fig:m2l2j-fat} at NLO and at {\tt POWHEG+PYTHIA} level. 
%
%
\FIGURE[t]{
\includegraphics[angle=0,width=0.65\textwidth]{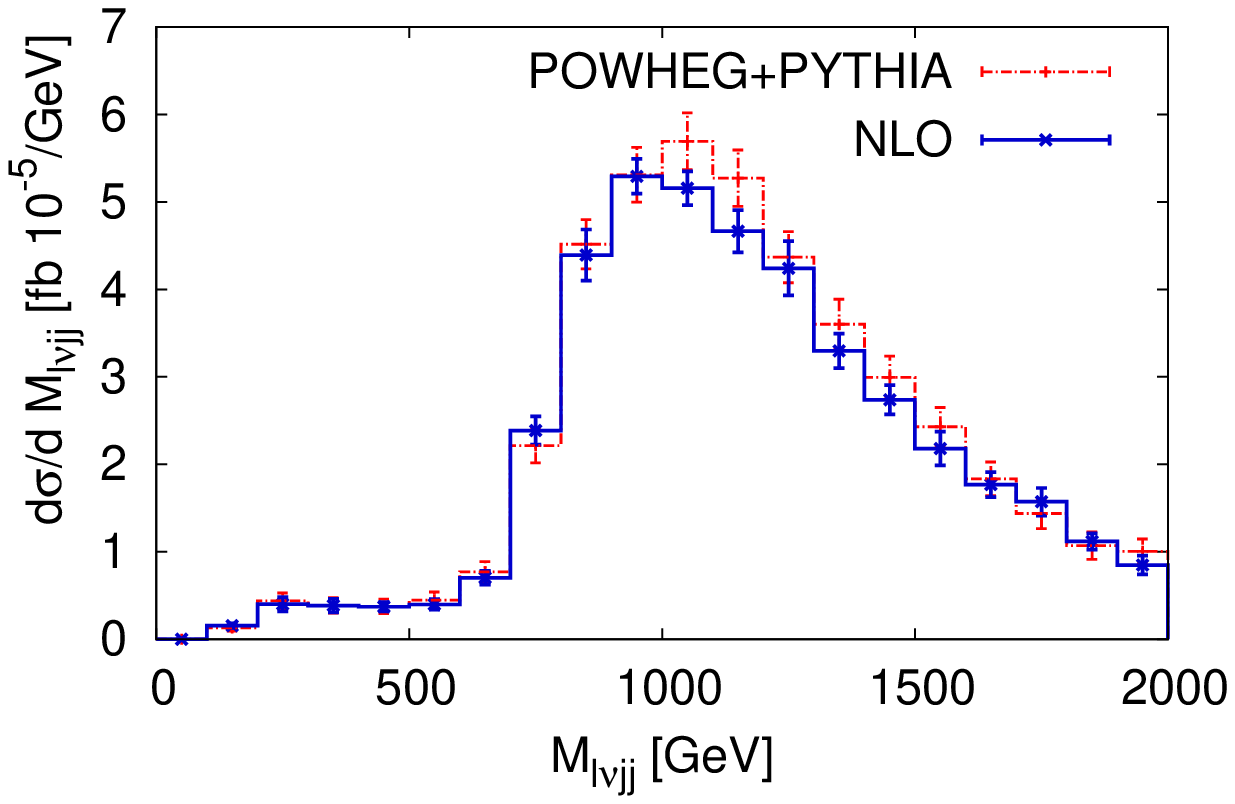}
  \caption{Invariant-mass distribution of the decay system in $pp\to
    W^+(jj) W^-(\mu^-\bar \nu_\mu)\,jj$ at the LHC with $\sqrt{s}=14$~TeV 
    for the boosted kinematics within the cuts of
    Eqs.~(\ref{eq:pt-boost})--(\ref{eq:gap-boost}), at NLO-QCD
    (blue solid lines) and with {\tt POWHEG+PYTHIA}~(red dashed lines), see text for details.}
\label{fig:m2l2j-fat}
} 
%
%
%
Because of the tight cuts imposed, this distribution peaks in the
1~TeV-region. As expected, the impact of the parton shower on the NLO results is
comfortably moderate, as apparent from the small differences between
the respective curves in the figure. The transverse momentum and
rapidity distributions of the tagging jets and the hard lepton are
similarly insensitive to parton-shower effects. 
%

\section{Conclusions}
\label{sec:conc}
In this work we have presented an implementation of electroweak
$\wpm\,jj$ production at hadron colliders in the {\tt POWHEG-BOX}, a
framework that allows the merging of NLO-QCD calculations with
transverse-momentum ordered parton shower programs such as {\tt
  PYTHIA}. We are providing code for fully leptonic and semi-leptonic
decay modes of the weak bosons, taking resonant and non-resonant
contributions and spin-correlations of the final-state particles into
account. Our implementation is publicly available from the {\tt
  POWHEG-BOX} webpage, {\tt http://powhegbox.mib.infn.it}.

While the user of the code is free to set parameters and analysis cuts
according to her own needs, we have presented phenomenological results
for three representative setups, which are of particular relevance for
understanding the mechanism of electroweak symmetry breaking realized
in nature. We have investigated VBF~$\wpm\,jj$ production with fully
leptonic decays of the weak bosons in the presence of a light Higgs
boson, and in a scenario with a heavy Higgs boson and semi-leptonic
decays of the weak bosons. In addition, we considered electroweak
$\wpm\,jj$ production in the kinematic regime where the weak bosons
are highly boosted and the hadronically decaying $W$~boson, ending up
in a fat jet, is identified via modern jet deconstruction techniques.

We found that the impact of the parton shower on most NLO
distributions is moderate in the fully leptonic decay mode, but some
care is needed in the definition of observables that are important for
central-jet veto techniques. When considering semi-leptonic decay
modes, however, significant differences arise between results at NLO
QCD and NLO results matched with {\tt PYTHIA}, unless boosted
techniques are used. These systematic
differences should be kept in mind when employing VBF~$\wpm\,jj$
production for precision studies at the LHC.

{\bf Acknowledgments} 
We are grateful to Andrea Banfi, Paolo Nason and Gavin Salam for
extensive, valuable discussions.  The work of B.~J.\ is supported by
the Research Center {\em Elementary Forces and Mathematical
  Foundations (EMG)} of the Johannes-Gutenberg-Universit\"at
Mainz. G.~Z.\ is supported by the British Science and Technology
Facilities Council, by the LHCPhenoNet network under the Grant
Agreement PITN-GA-2010-264564 and by the European Research and
Training Network (RTN) grant Unification in the LHC ERA under the
Agreement PITN-GA-2009-237920.

%


\begin{thebibliography}{99}

\bibitem{atlas:2012gk}
  G.~Aad {\it et al.}  [ATLAS Collaboration],
  Phys.\ Lett.\ B {\bf 716} (2012) 1
  [arXiv:1207.7214 [hep-ex]].


\bibitem{cms:2012gu}
  S.~Chatrchyan {\it et al.}  [CMS Collaboration],
  Phys.\ Lett.\ B {\bf 716} (2012) 30
  [arXiv:1207.7235 [hep-ex]].

\bibitem{Aaltonen:2012qt}
  T.~Aaltonen {\it et al.}  [CDF and D0 Collaborations],
  Phys.\ Rev.\ Lett.\  {\bf 109} (2012) 071804
  [arXiv:1207.6436 [hep-ex]].

\bibitem{Zeppenfeld:2000td}
  D.~Zeppenfeld, R.~Kinnunen, A.~Nikitenko and E.~Richter-Was,
  Phys.\ Rev.\ D {\bf 62} (2000) 013009
  [hep-ph/0002036].


\bibitem{Duhrssen:2004cv}
  M.~Duhrssen, S.~Heinemeyer, H.~Logan, D.~Rainwater, G.~Weiglein and D.~Zeppenfeld,
  Phys.\ Rev.\ D {\bf 70} (2004) 113009
  [hep-ph/0406323].

\bibitem{LHCHiggsCrossSectionWorkingGroup:2012nn}
  LHC Higgs Cross Section Working Group, A.~David, A.~Denner, M.~Duehrssen, M.~Grazzini, C.~Grojean, G.~Passarino and M.~Schumacher {\it et al.},
  arXiv:1209.0040 [hep-ph].

\bibitem{Rainwater:1999sd}
  D.~L.~Rainwater and D.~Zeppenfeld,
  Phys.\ Rev.\ D {\bf 60} (1999) 113004
   [Erratum-ibid.\ D {\bf 61} (2000) 099901]
  [hep-ph/9906218].

\bibitem{Kauer:2000hi}
  N.~Kauer, T.~Plehn, D.~L.~Rainwater and D.~Zeppenfeld,
  Phys.\ Lett.\ B {\bf 503} (2001) 113
  [hep-ph/0012351].

\bibitem{Rainwater:1997dg}
  D.~L.~Rainwater and D.~Zeppenfeld,
  JHEP {\bf 9712} (1997) 005
  [hep-ph/9712271].

\bibitem{Rainwater:1998kj}
  D.~L.~Rainwater, D.~Zeppenfeld and K.~Hagiwara,
  Phys.\ Rev.\ D {\bf 59} (1998) 014037
  [hep-ph/9808468].

\bibitem{Jager:2006zc}
  B.~J\"ager, C.~Oleari, D.~Zeppenfeld, 
  JHEP {\bf 07 } (2006)  015.
  [arXiv:hep-ph/0603177].

\bibitem{Jager:2006cp}
  B.~J\"ager, C.~Oleari, D.~Zeppenfeld, 
  Phys.\ Rev.\  {\bf D73 } (2006)  113006.
  [arXiv:hep-ph/0604200].


\bibitem{Bozzi:2007ur}
  G.~Bozzi, B.~J\"ager, C.~Oleari, D.~Zeppenfeld, 
  Phys.\ Rev.\  {\bf D75 } (2007)  073004.
  [arXiv:hep-ph/0701105].


\bibitem{Jager:2009xx}
  B.~J\"ager, C.~Oleari, D.~Zeppenfeld, 
  Phys.\ Rev.\  {\bf D80 } (2009)  034022.
  [arXiv:0907.0580 [hep-ph]].


\bibitem{Denner:2012dz}
  A.~Denner, L.~Hosekova and S.~Kallweit,
  arXiv:1209.2389 [hep-ph].

\bibitem{Jager:2011ms}
  B.~Jager and G.~Zanderighi,
  JHEP {\bf 1111} (2011) 055
  [arXiv:1108.0864 [hep-ph]].

\bibitem{Marchesini:1991ch}
G.~Marchesini {\it et al.}, 
Comp.\ Phys.\ Commun.\ {\bf 67} (1992) 465.
%
\bibitem{Corcella:2000bw}
  G.~Corcella {\it et al.},
  JHEP {\bf 0101 } (2001)  010.
  [hep-ph/0011363].

\bibitem{Sjostrand:2006za}
  T.~Sjostrand, S.~Mrenna, P.~Z.~Skands,
  JHEP {\bf 0605 } (2006)  026.
  [hep-ph/0603175].


\bibitem{Alioli:2010xd}
  S.~Alioli, P.~Nason, C.~Oleari, E. Re,
  JHEP {\bf 1006 } (2010)  043.
  [arXiv:1002.2581 [hep-ph]].


\bibitem{Nason:2004rx}
  P.~Nason,
  JHEP {\bf 0411 } (2004)  040.
  [hep-ph/0409146].

\bibitem{Frixione:2007vw}
  S.~Frixione, P.~Nason, C.~Oleari,
  JHEP {\bf 0711 } (2007)  070.
  [arXiv:0709.2092 [hep-ph]].


\bibitem{Arnold:2008rz}
K.~Arnold {\it et al.}, 
Comp.\ Phys.\ Comm.\ {\bf 180} (2009) 1661.
[arXiv:0811.4559 [hep-ph]]; \\
  K.~Arnold {\it et al.},
  arXiv:1107.4038 [hep-ph]; \\
  K.~Arnold  {\it et al.},
  arXiv:1207.4975 [hep-ph].


\bibitem{Ciccolini:2007ec}
  M.~Ciccolini, A.~Denner, S.~Dittmaier,
  Phys.\ Rev.\  {\bf D77 } (2008)  013002.
  [arXiv:0710.4749 [hep-ph]].

\bibitem{Groote:2013xt}
  S.~Groote, J.~G.~Korner and P.~Tuvike,
  arXiv:1301.0881 [hep-ph].


\bibitem{Denner:2002ii}
A.~Denner, S.~Dittmaier, 
Nucl.\ Phys.\ {\bf B658} (2003) 175.
[hep-ph/0212259].

\bibitem{Denner:2005nn}
A.~Denner, S.~Dittmaier, 
Nucl.\ Phys.\ {\bf B734} (2006) 62.
[hep-ph/0509141].

\bibitem{Frixione:1995ms}
S.~Frixione, Z.~Kunszt, A.~Signer,
Nucl.\ Phys.\ {\bf B467} (1996) 399. 
[hep-ph/9512328]. 

\bibitem{Jager:2012xk}
  B.~Jager, S.~Schneider, G.~Zanderighi,
  JHEP {\bf 1209} (2012) 083
  [arXiv:1207.2626 [hep-ph]].


\bibitem{Nason:2009ai}
  P.~Nason, C.~Oleari,
  JHEP {\bf 1002 } (2010)  037.
  [arXiv:0911.5299 [hep-ph]].

\bibitem{Cacciari:2005hq}
M.~Cacciari, G.~P.~Salam, 
Phys.\ Lett.\ {\bf B641} (2006) 57. 
[hep-ph/0512210]. 

\bibitem{Cacciari:2008gp}
  M.~Cacciari, G.~P.~Salam and G.~Soyez,
  JHEP {\bf 0804} (2008) 063
  [arXiv:0802.1189 [hep-ph]].

\bibitem{Dittmar:1996ss}
  M.~Dittmar and H.~K.~Dreiner,
  Phys.\ Rev.\ D {\bf 55} (1997) 167
  [hep-ph/9608317].

\bibitem{Stelzer:1994ta}
  T.~Stelzer and W.~F.~Long,
  Comput.\ Phys.\ Commun.\  {\bf 81} (1994) 357
  [hep-ph/9401258].

\bibitem{Alwall:2007st}
J.~Alwall {\it et al.},
  JHEP {\bf 0709} (2007) 028
  [arXiv:0706.2334 [hep-ph]].


\bibitem{Martin:2009iq}
  A.~D.~Martin, W.~J.~Stirling, R.~S.~Thorne, G.~Watt,
  Eur.\ Phys.\ J.\  {\bf C63 } (2009)  189-285.
  [arXiv:0901.0002 [hep-ph]].

\bibitem{Whalley:2005nh}
M.~R.~Whalley, D.~Bourilkov, R.~C.~Group, 
hep-ph/0508110. 


\bibitem{Cacciari:2011ma}
  M.~Cacciari, G.~P.~Salam and G.~Soyez,
  Eur.\ Phys.\ J.\ C {\bf 72} (2012) 1896
  [arXiv:1111.6097 [hep-ph]].


\bibitem{ATLAS-HWW-2012}
ATLAS Collaboration, 
{\tt http://cdsweb.cern.ch/record/1462530/files/ATLAS-CONF-2012-098}, 
ATLAS-CONF-2012-098 (2012). 

\bibitem{Hamilton:2012np}
  K.~Hamilton, P.~Nason and G.~Zanderighi,
  JHEP {\bf 1210} (2012) 155
  [arXiv:1206.3572 [hep-ph]].


\bibitem{Barger:1994zq}
  V.~D.~Barger, R.~J.~N.~Phillips and D.~Zeppenfeld,
  Phys.\ Lett.\ B {\bf 346} (1995) 106
  [hep-ph/9412276].


\bibitem{Aad:2012me}
  G.~Aad {\it et al.}  [ATLAS Collaboration],
  arXiv:1206.6074 [hep-ex].

\bibitem{Abdesselam:2010pt}
  A.~Abdesselam {\it et al.},
  Eur.\ Phys.\ J.\ C {\bf 71} (2011) 1661
  [arXiv:1012.5412 [hep-ph]].

\bibitem{Butterworth:2002tt}
  J.~M.~Butterworth, B.~E.~Cox and J.~R.~Forshaw,
  Phys.\ Rev.\ D {\bf 65} (2002) 096014
  [hep-ph/0201098].

\bibitem{Butterworth:2008iy}
  J.~M.~Butterworth, A.~R.~Davison, M.~Rubin and G.~P.~Salam,
  Phys.\ Rev.\ Lett.\  {\bf 100} (2008) 242001
  [arXiv:0802.2470 [hep-ph]].

\bibitem{Dokshitzer:1997in}
  Y.~L.~Dokshitzer, G.~D.~Leder, S.~Moretti and B.~R.~Webber,
  JHEP {\bf 9708} (1997) 001
  [hep-ph/9707323].

\bibitem{Wobisch:1998wt}
  M.~Wobisch and T.~Wengler,
  In *Hamburg 1998/1999, Monte Carlo generators for HERA physics* 270-279
  [hep-ph/9907280].




\end{thebibliography}
\end{document}